\documentstyle[aps,twocolumn]{revtex}

\begin{document}
\title{Electric Field Effect in Ultrathin Films near the Superconductor-Insulator
Transition }
\author{N. Markovi\'{c}\thanks{%
Present address: Department of Physics, Harvard University, Cambridge, MA},
C. Christiansen, G. Martinez-Arizala, and A. M. Goldman}
\address{School of Physics and Astronomy, University of Minnesota, \\
Minneapolis,\\
MN 55455, USA}
\date{May 27, 2001}
\maketitle

\begin{abstract}
The effect of an electric field on the conductance of ultrathin films of
metals deposited on substrates coated with a thin layer of amorphous Ge was
investigated. A contribution to the conductance modulation symmetric with
respect to the polarity of the applied electric field was found in regimes
in which there was no sign of glassy behavior. For films with thicknesses
that put them on the insulating side of the superconductor-insulator
transition, the conductance increased with electric field, whereas for films
that were becoming superconducting it decreased. Application of magnetic
fields to the latter, which reduce the transition temperature and ultimately
quench superconductivity, changed the sign of the reponse of the conductance
to electric field back to that found for insulators. We propose that this
symmetric response to capacitive charging is a consequence of changes in the
conductance of the {\it a-}Ge layer, and is not a fundamental property of
the physics of the superconductor-insulator transition as previously
suggested.
\end{abstract}


Investigations of the effect of a perpendicular applied electric field on
insulating granular Au films \cite{Adkins} and amorphous $In_{2}O_{3}$ films 
\cite{Ben-Chorin} revealed that the conductance increased for both
polarities of field. This symmetric response was ascribed to nonequilibrium
transport phenomena specific to very disordered systems. Long relaxation
times and memory , aging and hysteresis associated with the symmetric
response support the picture that these systems are Coulomb glasses with
long equilibrium times \cite{Davies,Bhatt}. More recently, the inhomogeneous
nature of charge transport and the slow relaxation in such systems has been
studied theoretically \cite{Yu,Pastor} and experimentally as a function of
disorder and magnetic field \cite{Ovadyahu,Vaknin}.

A similar symmetric response to an applied electric field was found in
ultrathin amorphous films of Bi and Pb \cite{Martinez1,Martinez2} grown on
amorphous Ge ({\it a}-Ge). We speculated that in such very resistive films
the variation of the conductance with electric field was proportional to the
dependence of the electronic density of states on energy, since the excited
state caused by the applied field cannot relax in very glassy systems \cite
{Martinez2}. This conjecture has been substantiated by recent simulations of
the time development of the Coulomb gap in a Coulomb glass \cite{Yu}. There
are several problematic aspects to this interpretation: in ultrathin films
of Bi and Pb, the symmetric response persisted into regimes where the glassy
behavior vanishes, such as at high temperatures and in more conductive
(thicker) films. Furthermore, when the films were thick enough to become
superconducting, the response changed sign and the conductance above the
transition temperature {\it decreased} as a function of the applied electric
field \cite{Martinez1}. The largest fractional conductance modulation was
observed deep in the insulating regime, and it decreased as film thickness
increased and the insulator-to-superconductor transition was approached. At
the transition, the symmetry disappeared, and for a small range of
thicknesses around the transition, the field effect was approximately
antisymmetric, before changing sign in films that were thick enough to be
superconducting at low temperatures. The carrier density in these studies
could not be increased enough to drive a nonsuperconducting film into the
superconducting state, almost certainly a consequence of the high density of
trapping sites. A qualitative interpretation of these results that was
presented, included consideration of the possibility of Cooper pairing in
insulating films and symmetry between the insulating and the superconducting
states.

Here we report on investigations of the electric field effect in ultrathin
films of Bi at temperatures which traverse a range of temperatures an order
of magnitude lower in temperature than those of our previous work. In
addition, the superconductor-insulator transition was tuned not only by
changing film thickness, but by the application of a perpendicular magnetic
field. Films thick enough to become superconducting and exhibiting a
conductance above their transition that decreased with applied electric
field, were made to revert to the behavior of insulating films whose
conductance decreased with electric field. This ensued on application of a
perpendicular magnetic field. Although these results are a consistent
extension of earlier work, they lead to a new interpretation. We now propose
that the symmetric response to an electric field for both superconducting
and insulating films is due to the conductivity of the amorphous Ge layer
being increased by capacitive charging, rather than any underlying symmetry
between the insulating and superconducting states.

These investigations were carried out on a series of ultrathin Bi films,
evaporated on top of a $10\AA $\ thick layer of {\it a-}Ge, which was
pre-deposited onto a $SrTiO_{3}$ $(100)$ substrate. The substrate
temperature was kept below $20K$ during all depositions, and all of the
films were grown{\it \ in situ} under UHV conditions ($\sim $ $10^{-10}$
Torr). The film thickness was gradually increased through successive
depositions in increments of $0.1-0.2\AA $. Films grown in this manner are
believed to be homogeneous, since they become connected at an average
thickness of about one monolayer \cite{Strongin}.

For capacitive charging studies, the electric field was applied
perpendicular to the plane by biasing the film relative to a $100nm$ thick
Au metal gate on the back of the $SrTiO_{3}$ substrate, which was $0.75mm$
thick. Even though these substrates are of macroscopic thickness, a
substantial charge can be induced at relatively low gate voltages because of
their high dielectric constant below $10K$ $(\kappa \varepsilon _{0}\sim
8000-20000)$ \cite{Christen}.

Resistance measurements were carried out between depositions using a
standard four-probe DC technique. Low bias currents $(<50nA)$ were used to
avoid Joule heating of the sample and to make sure that the voltage across
the sample was a linear function of the applied current. As the film
thickness increased from $7\AA $\ to $15\AA $, the temperature dependence of
the resistance of the system changed from insulator-like ($dR/dT<0$) to
superconductor-like ($dR/dT>0$) at low temperatures \cite{Markovic1}. There
was no sign of the quasi-reentrant behavior observed in granular films \cite
{Jaeger}. In order to study the electric field response of a film across the
magnetic field driven superconductor-insulator transition, we focussed on
films which were superconducting in zero field, but could be driven
insulating by applying a magnetic field perpendicular to the plane of the
sample \cite{Markovic2}.

The sheet resistance of a representative film as a function of temperature
at different magnetic fields is shown in Fig. \ref{Fig2}. In zero field, the
resistance decreases as the temperature decreases, indicating the onset of
superconductivity. At the lowest temperatures, as the magnetic field is
gradually increased, the temperature coefficient of the resistance, $dR/dT,$
changes sign, indicating insulating behavior.

The change in the conductance with gate voltage in zero magnetic field of
the same film at four temperatures is shown in Fig. \ref{Fig3}. It is
evident that the conductance decreases with applied gate voltage at
temperatures for which $dR/dT>0$ (compare to the zero-field curve in Fig. 
\ref{Fig2}.), and the effect becomes smaller at higher temperatures. This is
similar to the result Martinez-Arizala {\it et al}. \cite{Martinez1},
obtained for superconducting Pb films. No relaxation effects were observed
when gate voltage was applied in this regime.

The change in conductance as a function of gate voltage at a temperature of $%
0.15K$ in different magnetic fields is shown on Fig. \ref{Fig4}. The
conductance is found to decrease for both polarities of the gate voltage in
low fields, even though the effect is not perfectly symmetric. At some value
of the magnetic field, where $dR/dT$ is still positive, but small, the
effect becomes approximately antisymmetric. The data for positive polarity
of the gate voltage look remarkably similar to the data shown in Fig. \ref
{Fig3}., even though the sample was driven into the insulating state by
increasing the magnetic field in one case, and by increasing the temperature
in the other.

In a magnetic field which is high enough to bring the sample into the $%
dR/dT<0$ regime, the conductance increases as the gate voltage is applied,
as shown on Fig. \ref{Fig5}. Even though the effect is relatively small, it
is reproducible. The results are remarkably similar to the case when the
superconductor-insulator transition is tuned by changing the film thickness 
\cite{Martinez1}. Furthermore, similar results are obtained if the sample is
driven out of the superconducting state by increasing the temperature. The
sign and value of $dR/dT$ seem to indicate whether the field effect will be
positive or negative, and whether it will be symmetric or antisymmetric.

We now address the physical nature of the capacitive charging experiment. In
particular we will consider the issue of why the conductance-gate voltage
characteristic resembles the density of states of a disordered system. The
Efros-Shklovskii density of states (DOS) for two dimensional disordered
systems is a vee-shaped entity with its apex centered on the Fermi energy ($%
E_{F}$) \cite{Shklovskii,Levin,Sarvestani}. Conductance studies do not
usually probe the DOS of electronic systems, since disturbances to the
electron energy distribution relax, and carriers added during a measurement
are quickly screened \cite{Monroe}. This does not occur in the case of
tunneling, as the physical process, characterized by the tunneling time, is
short in comparison with the time of charge relaxation. As a consequence,
tunneling is the standard approach to the measurement of the density of
states \cite{Massey,McMillan,Hertel,White1,White2,Fritzshe}.

The usual situation in a capacitive charging experiment is one in which the
electron distribution relaxes after the gate voltage and carrier
concentration are changed. The new carrier concentration determines the
value of the chemical potential, and the minimum of the DOS tracks that new
value \cite{Shklovskii}. We pointed out earlier that in glassy systems with
long relaxation times, there may not be a path from the excited state
resulting from injected carriers back to the state of minimum free energy,
because of barriers in the free energy landscape. Thus changing the gate
voltage would change $E_{F}$, but the minimum of the DOS would not track it.
In this circumstance, if the conductance were proportional to the density of
states, as is the case in the hopping regime \cite{Kamimura}, a simple field
effect conductance modulation study could map out the dependence of the DOS
on energy. Indeed, recent simulations have shown that the time development
of the Coulomb gap in the DOS can involve very long time scales due to
electron hopping and rearrangement \cite{Yu}.

In the case of the ultrathin films of metals deposited on top of a thin {\it %
a-}Ge layer grown on a $SrTiO_{3}$ substrate, the geometry is very critical
to understanding the evolution of the response to capacitive charging. For
the very thinnest films, the order of a $10\AA $ thick layer of metal on top
of a $6\AA $ to $10\AA $ thick layer of {\it a-}Ge, the system behaves as an
electron glass, and the conductance modulation, which can be a significant
fraction of the total conductance, is a direct measure of the DOS by the
argument given above. The relevant density of states is in effect that of
the {\it a-}Ge layer, with the metal layer acting as a source of electrons.
In fact, tunneling measurements of the electronic density of states of
amorphous $Ge_{1-x}Au_{x}$ show a very similar shape as a function of the
gate voltage as that found in our experiments \cite{McMillan}. The response
is symmetric in voltage because of the symmetry of the DOS for a disordered
system.

As film thickness is increased, the glass-like response of the conductance
disappears, but there is a small conductance change that still exhibits a
vee shape, characteristic of the DOS of a disordered system. This behavior
can be understood if one appreciates the continued presence of the {\it a-}
Ge underlayer, in which charge can be confined. When the gate voltage is
increased, charges are drawn out of the electrodes connected to the film,
and the metal film's carrier concentration is changed by a small amount. The
gate voltage and the effective capacitance to the gate determine the amount
of charge in the {\it a-}Ge layer, which determines the chemical potential.
The metal layer permits a rapid change of the chemical potential, but
internal to the {\it a-}Ge the charges have a long relaxation time, because
the {\it a-}Ge is extremely disordered and separated from the metal film by
a Schottky barrier \cite{Schottky}. By the arguments given above, this
results in changes in the conductance of that layer proportional to its DOS.
The total fraction of the modulation of the conductance attributable to the 
{\it a-}Ge layer is only a small fraction of the total conductance, so the
overall effect is small.

When the metal film becomes superconducting other physics comes into play.
Increasing or decreasing the gate voltage from zero increases the
conductance of the {\it a-}Ge layer. The combination of the superconducting
layer and the {\it a-}Ge layer can be considered to be a proximity sandwich.
Increasing the conductance of the {\it a-}Ge layer will make it more
metallic, causing the electrons in the sandwich to have a higher probability
of being in the {\it a-}Ge layer. Thus the average attractive
electron-electron interaction which they experience will be reduced,
lowering the effective transition temperature of the sandwich. The proximity
of a metallic layer has been shown to reduce the superconducting transition
temperature \cite{Shapovalov} in some cases. At fixed temperature, in the
regime where the resistance is decreasing with decreasing temperature, a
reduced transition temperature derived from capacitive charging, will result
in an increase of the resistance with increasing gate voltage (of either
sign). Thus the effect of initiating the superconducting transition in the
overlay film is to flip over the vee-shaped response to the gate voltage.
Again, because the {\it a-}Ge layer at this point contributes only a small
part of the total conductance of the composite film, the effect is small.
Correspondingly, quenching superconductivity with a magnetic field, flips
the conductance-voltage characteristic back to its behavior in the
insulating regime.

It should be noted that the original goal of these studies, which was not
realized, was to induce superconductivity through changing the carrier
concentration by capacitive charging. Superconductivity has been induced in
a similar experimental configuration by Sch\"{o}n {\it et. al.} in organic
crystals\cite{Schon}, but at much higher gate voltages than were used in our
studies of ultrathin metal films. It remains the subject of future
investigations as to whether superconductivity can be induced in ultrathin
metal films deposited on {\it a-}Ge by capacitive charging, if the gate
voltage were large enough to transfer sufficient charge to both fill all of
the traps in the {\it a-}Ge and change the carrier concentration in the
metal layer.

In summary, we have performed the field-effect conductance modulation
experiment on ultrathin films of Bi deposited onto amorphous Ge near the
thickness- and magnetic field-tuned superconductor-insulator transitions. We
have swept through the superconductor-insulator transition using a magnetic
field, and we see the same change of sign in the field effect response
observed when the superconductor-insulator transition is traversed by
changing thickness. The results of these studies together with our previous
work on both Pb and Bi \cite{Martinez1,Martinez2}, lead us to propose that
the small symmetric response in this regime, where no glassy behavior was
observed, is due to increasing the conductivity of the insulating amorphous
germanium underlayer, rather than some symmetry between the insulating and
the superconducting states.

We gratefully acknowledge useful discussions with B. Shklovskii and P.
Allen. This work was supported in part by the National Science Foundation
under Grant No. NSF/DMR-9876816. 

\begin{figure}[tbp]
\caption{Sheet resistance as a function of temperature for a 14\AA\ thick Bi
film in different magnetic fields: 0, 1, 2, 3, 4, 5, 8, 11 and 12 kG, from
bottom to top.}
\label{Fig2}
\end{figure}
\begin{figure}[tbp]
\caption{Fractional change in conductance of a 14\AA\ thick Bi film as a
function of gate voltage at 0.15 K (triangles), 0.2 K (diamonds), 0.3 K
(squares) and 0.4 K (circles).}
\label{Fig3}
\end{figure}
\begin{figure}[tbp]
\caption{Absolute change in conductance of a 14\AA\ thick Bi film as a
function of gate voltage at 0.15 K in zero magnetic field (circles), and in
a perpendicular magnetic field of 1 kG (squares), 2 kG (diamonds), 3 kG
(crosses) and 4 kG (triangles).}
\label{Fig4}
\end{figure}
\begin{figure}[tbp]
\caption{Absolute change in conductance of a 14\AA\ thick Bi film as a
function of gate voltage at 0.15 K in a perpendicular magnetic field ranging
from 1 kG (bottom) up to 12 kG (top).}
\label{Fig5}
\end{figure}



\begin{references}
\bibitem{Adkins}  C. J. Adkins, J. D. Benjamin, J. M. D. Thomas, J. W.
Gardner, and A. J. McGeown, J. Phys. C {\bf 17}, 4633 (1984).

\bibitem{Ben-Chorin}  M. Ben-Chorin, D. Kowal, and Z. Ovadyahu, Phys. Rev. B 
{\bf 44}, 3420 (1991); M. Ben-Chorin, Z. Ovadyahu and M. Pollak, Phys. Rev.
B {\bf 48}, 15 025 (1993).

\bibitem{Davies}  J. H. Davies, P. A. Lee, and T. M. Rice, Phys. Rev. B {\bf %
29}, 4260 (1984).

\bibitem{Bhatt}  R. N. Bhatt and T. V. Ramakrishnan, J. Phys. C {\bf 17},
L639 (1984).

\bibitem{Yu}  Clare C.Yu, Phys. Rev. Lett. {\bf 82}, 4074 (1999).

\bibitem{Pastor}  A. A. Pastor and V. Dobrosavljevi\'{c}, Phys. Rev. Lett. 
{\bf 83}, 4642 (1999).

\bibitem{Ovadyahu}  Z. Ovadyahu and M. Pollak, Phys. Rev. Lett. {\bf 79},
459 (1997).

\bibitem{Vaknin}  A. Vaknin, Z. Ovadyahu, and M. Pollak, Phys. Rev. Lett. 
{\bf 81}, 669 (1998); Phys. Rev. Lett. {\bf 84}, 3402 (2000); Phys. Rev. B 
{\bf 61}, 6692 (2000).

\bibitem{Martinez1}  G. Martinez-Arizala, D. E. Grupp, C. Christansen,
A.M.Mack, N. Markovi\'{c}, Y. Seguchi, and A. M. Goldman, Phys. Rev. Lett. 
{\bf 78}, 1130 (1997).

\bibitem{Martinez2}  G. Martinez-Arizala, C. Christansen, D. E. Grupp, N.
Markovi\'{c}, A.M.Mack, and A. M. Goldman, Phys. Rev. B {\bf 57}, R670
(1998).

\bibitem{Strongin}  M. Strongin, R. S. Thompson, O. F. Kammerer,and J. E.
Crow, Phys. Rev. B {\bf 1}, 1078 (1970).

\bibitem{Christen}  H. M. Christen, J. Manhart, E. J. Williams, and Ch.
Gerber, Phys. Rev. B {\bf 49}, 12 095 (1994).

\bibitem{Markovic1}  N. Markovi\'{c}, C Christiansen, and A. M. Goldman,
Phys. Rev. Lett. {\bf 81}, 5217 (1998).

\bibitem{Jaeger}  H. M. Jaeger, D. B. Haviland, B. G. Orr, and A. M.
Goldman, Phys. Rev. B {\bf 40}, 182 (1989).

\bibitem{Markovic2}  N. Markovi\'{c}, C Christiansen, A. M. Mack, W. H.
Huber, and A. M. Goldman, Phys. Rev. B {\bf 60}, 4320 (1999).

\bibitem{Shklovskii}  B. I. Shklovskii and A. L. Efros, {\it Electronic
Properties of Doped Semiconductors} (Springer, Berlin, 1984).

\bibitem{Levin}  E. I. Levin, V. L. Nguen, B. I. Shklovski, and A. L. Efros,
Zh. Exp. Teor. Fiz. {\bf 92}, 1499 (1987) [Sov. Phys. JETP {\bf 65}, 842
(1987)].

\bibitem{Sarvestani}  M. Sarvestani, M. Schreiver, and T. Bojta, Phys. Rev.
B {\bf 52}, R3850 (1995).

\bibitem{Monroe}  D. Monroe, A. C. Gossard, J. H. English, B. Golding, W. H.
Haemmerle, and M. A. Kastner, Phys. Rev. Lett. {\bf 59}, 1148 (1987).

\bibitem{Massey}  J. G. Massey and Mark Lee, Phys. Rev. Lett. {\bf 75}, 4266
(1995).

\bibitem{McMillan}  W. L. McMillan and J. Mochel, Phys. Rev. Lett. {\bf 46},
556 (1981).

\bibitem{Hertel}  G. Hertel et al., Phys. Rev. Lett. {\bf 50}, 743 (1983).

\bibitem{White1}  A. E. White, R. C. Dynes, and J. P. Garno, Phys. Rev.
Lett. 56, 532 (1985).

\bibitem{White2}  A. E. White, R. C. Dynes, and J. P. Garno, Phys. Rev. B
31, 1174 (1986).

\bibitem{Fritzshe}  H. Fritzshe, in {\it Electronic and Structural
Properties of Amorphous Semiconductors} (Academic Press, New York, 1973), p
55.

\bibitem{Kamimura}  H. Kamimura and H. Aoki, {\it The Physics of Interacting
Electrons in Disordered Systems} (Oxford Science Publications, Clarendon
Press, Oxford, 1989), p. 123.

\bibitem{Schottky}  By depositing very small amounts of metal on top of the 
{\it a-}Ge layer, we may in fact be doping the Ge, and the observed
conductance is partially due to the doped Ge film. When the Bi film on top
actually becomes metallic, it shunts the Ge film, and a Schottky barrier
forms at the interface.

\bibitem{Shapovalov}  D. L. Shapovalov, Pis'ma Zh. Eksp. Teor. Fiz. {\bf 60}
, 193 (1994) [JETP Lett. {\bf 60}, 199 (1994)].

\bibitem{Schon}  J. H. Sch\"{o}n, Ch. Kloc, R.C. Haddon, and B. Batlogg,
Science {\bf 288}, 656 (2000). J. H. Sch\"{o}n, Ch. Kloc, and B. Batlogg,
Nature {\bf 406}, 702 (2000).
\end{references}
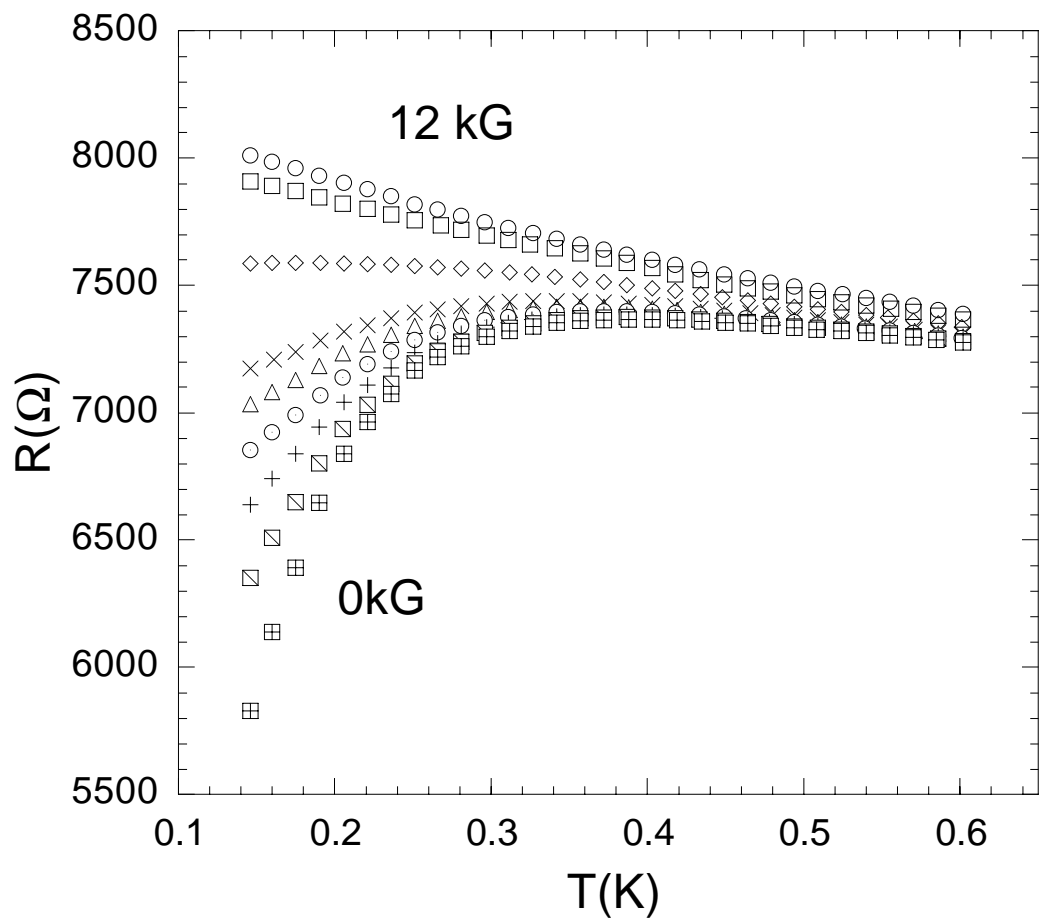
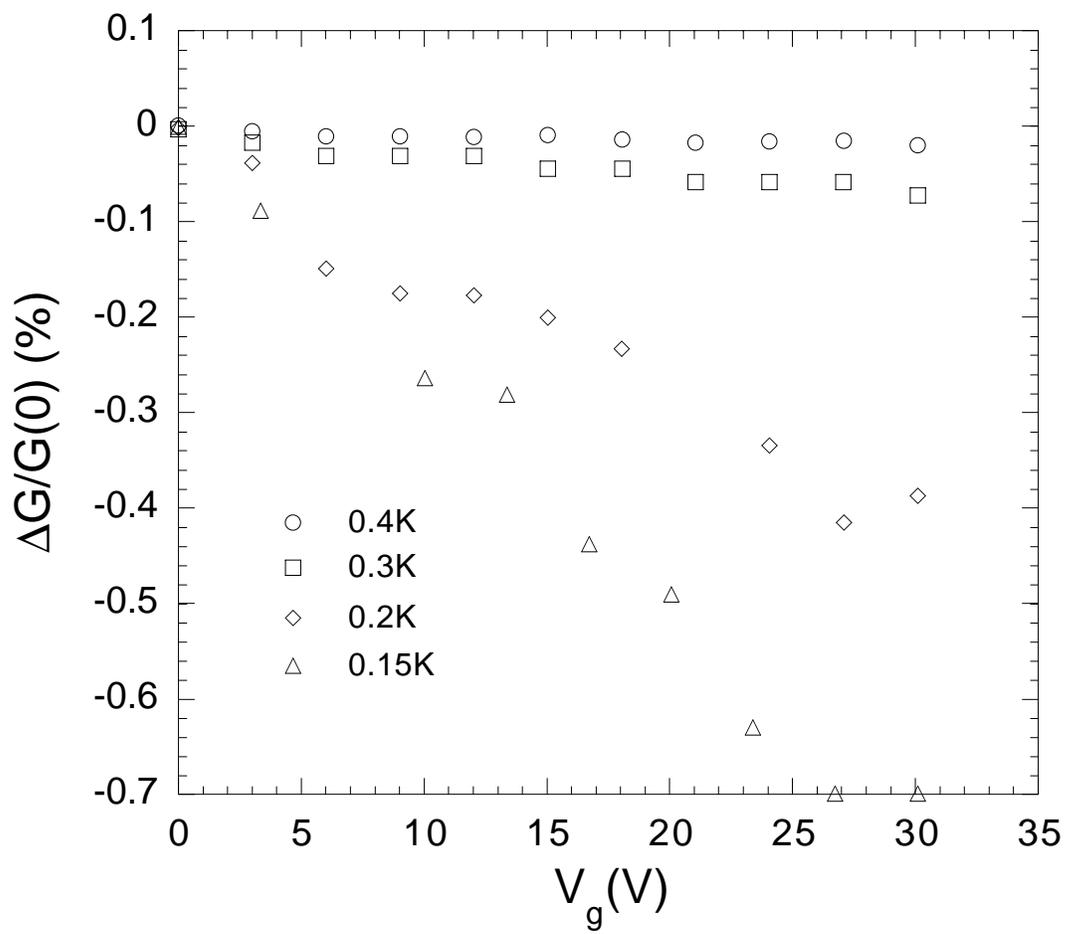
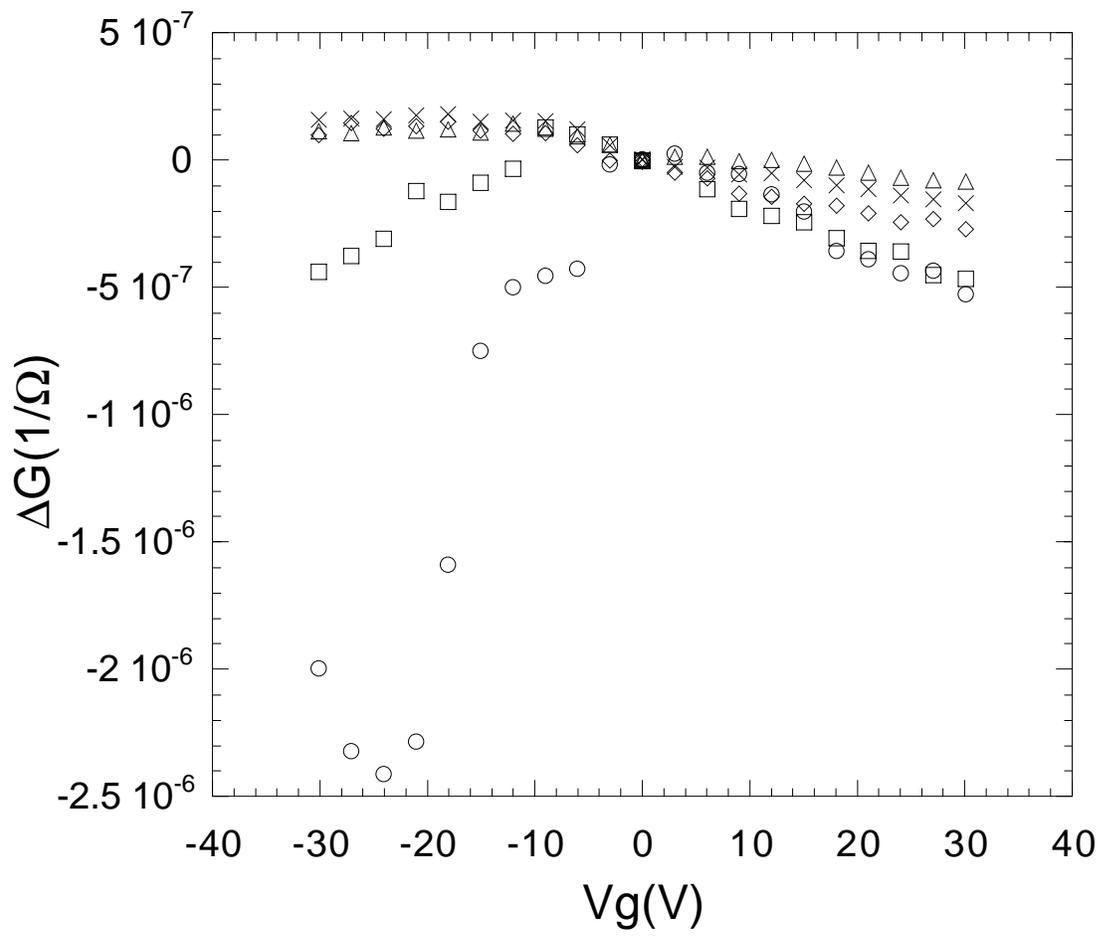
\end{document}